\documentclass[10pt,conference]{IEEEtran}

\usepackage[T1]{fontenc}
\usepackage{graphicx}
\usepackage{amsmath}
\usepackage{newtxmath}
\usepackage{url}
\usepackage{multirow}
\usepackage[nobreak]{cite}
\usepackage{xspace}
\usepackage{multicol}
\usepackage{booktabs}
\usepackage[table]{xcolor}
\usepackage{pdfpages}

\newcommand{\RQ}[1]{RQ${}_{\text{#1}}$\xspace}
\def\etal{\textit{et al.}\xspace}

\def\TP{\mathit{TP}}
\def\FN{\mathit{FN}}
\def\FP{\mathit{FP}}

\def\To{\Rightarrow}
\def\Late{${}^*$}
\def\Impl{${}^\diamond$}

\newcommand{\Pair}[2]{\langle\text{#1},\text{#2}\rangle}

\newcommand{\Smell}[1]{\textsf{#1}}
\newcommand{\Metric}[1]{\textsf{#1}}
\newcommand{\Predicate}[1]{\textit{#1}}
\newcommand{\U}[1]{\underline{#1}}

\newcommand{\Characteristic}[1]{\textit{#1}}
\def\Ambiguity{\Characteristic{Ambiguity}\xspace}
\def\Incorrectness{\Characteristic{Incorrectness}\xspace}
\def\Granularity{\Characteristic{Granularity}\xspace}
\def\Redundancy{\Characteristic{Redundancy}\xspace}
\def\Lack{\Characteristic{Lack}\xspace}
\def\Misplacement{\Characteristic{Misplacement}\xspace}
\def\Inconsistency{\Characteristic{Inconsistency}\xspace}

\newcommand{\Scope}[1]{\textit{#1}}
\def\Usecase{\Scope{Usecase}\xspace}
\def\Section{\Scope{Section}\xspace}
\def\Flow{\Scope{Flow}\xspace}
\def\Sentence{\Scope{Sentence}\xspace}
\def\Word{\Scope{Word}\xspace}

\begin{document}
\title{Detecting Bad Smells in Use Case Descriptions}

\author{%
  \IEEEauthorblockN{Yotaro Seki, Shinpei Hayashi, and Motoshi Saeki}
  \IEEEauthorblockA{%
    Tokyo Institute of Technology, Tokyo 152--8550, Japan\\
    Email: \{yotaro,hayashi,saeki\}@se.cs.titech.ac.jp}
}

\maketitle
\begin{abstract}
Use case modeling is very popular to represent the functionality of the system to be developed, and it consists of two parts: use case diagram and use case description.
Use case descriptions are written in structured natural language~(NL), and the usage of NL can lead to poor descriptions such as ambiguous, inconsistent and/or incomplete descriptions, etc.
Poor descriptions lead to missing requirements and eliciting incorrect requirements as well as less comprehensiveness of produced use case models.
This paper proposes a technique to automate detecting bad smells of use case descriptions, symptoms of poor descriptions.
At first, to clarify bad smells, we analyzed existing use case models to discover poor use case descriptions concretely and developed the list of bad smells, i.e., a catalogue of bad smells.
Some of the bad smells can be refined into measures using the Goal-Question-Metric paradigm to automate their detection.
The main contribution of this paper is the automated detection of bad smells.
We have implemented an automated smell detector for 22 bad smells at first and assessed its usefulness by an experiment.
As a result, the first version of our tool got a precision ratio of 0.591 and recall ratio of 0.981.
\end{abstract}

\section{Introduction}\label{s:introduction}

Use case modeling is one of the popular techniques to elicit requirements to business processes, information systems, and software~(simply, software hereafter), and is being made into practice \cite{ApplyingUseCases}.
Software developers can apply use case models to validate software design to requirements, to make test plans and development schedules, etc., and use case modeling can have wide varieties of applications.

A use case model consists of two parts: use case diagram and use case description.
Use case descriptions are written in structured natural language~(NL), and the usage of NL can lead to poor descriptions such as ambiguous, inconsistent and/or incomplete descriptions, etc.
Poor descriptions lead to missing requirements and eliciting incorrect requirements as well as less comprehensiveness of produced use case models.
In fact, many use case models of lower quality have been produced \cite{el-attar-sosym2010}.
Thus, it is significant to detect poor use case descriptions during the development of a use case model.

There have been developed several guidelines to compose use case descriptions of higher quality and checklists to detect problematic parts in them.
For example, Phalp \etal developed a checklist called 7C's which was specified in natural language \cite{phalp-sqj2007}, and T\"{o}rner \etal provided questionnaires as their quality checklists \cite{torner-isese2006}.
However, these guidelines can be applied manually only, and they allow us to miss problematic parts in use case descriptions.
Manual decision making of a poor description is subjective, and the results may depend on the requirements analysts.
Furthermore, manual checking is a time-consuming task for the analysts \cite{phalp-sqj2007}.

This paper proposes a technique to automate detecting ``bad smells'' of use case descriptions, symptoms of poor descriptions.
First of all, for use case descriptions, we define ``bad smells'', where the concept of bad smells has been established mainly for source code as their surface characteristics that might cause deep problems \cite{refactoring}.
We make a catalogue of bad smells from two views: their characteristics and levels of granularity~(scope) of occurring bad smells.
Note that bad smells are mainly not \emph{bugs} but indicators of poor descriptions that may cause some serious problems in future steps of a software development project.
To make the catalogue, we have collected 30 use case descriptions of various problem domains by using the Internet or other resources such as textbooks and tried to find ``bad smells'' out of them.
As a result, we got 54 bad smells.
Next, by applying Goal-Question-Metric~(GQM) paradigm \cite{GQM}, we got metrics to detect 22 of the 54 bad smells and developed an automated tool based on these metrics.
Though evaluating our catalogue and the automated tool, we found additional six bad smells and two metrics.
Finally, we got the precision ratio of 0.596 and recall ratio of 1.00 to detect bad smells by our final version of the automated tool.
The main contribution of this paper is the automated detection of bad smells.
Although the obtained catalogue of bad smells is a by-product, it is useful not only as a checklist to detect bad smells manually but also as a starting point to design the automated tool.

The rest of this paper is organized as follows.
We will present our analysis method and the developed catalogue of bad smells in the next section.
We also show the application of the GQM approach and the set of the derived metrics in the section.
Section~\ref{s:evaluation} presents the experimental evaluations of the catalogue and the automated tool.
In this section, we used the other eight use case descriptions different from the 30 descriptions used for developing the catalogue and the metrics.
Sections~\ref{s:relatedwork} and \ref{s:conclusion} are for related work and concluding remarks.

\section{Our Approach}\label{s:approach}

\subsection{Overview}\label{ss:overview}

\begin{figure}[bt]\centering
  \includegraphics[width=0.6\linewidth]{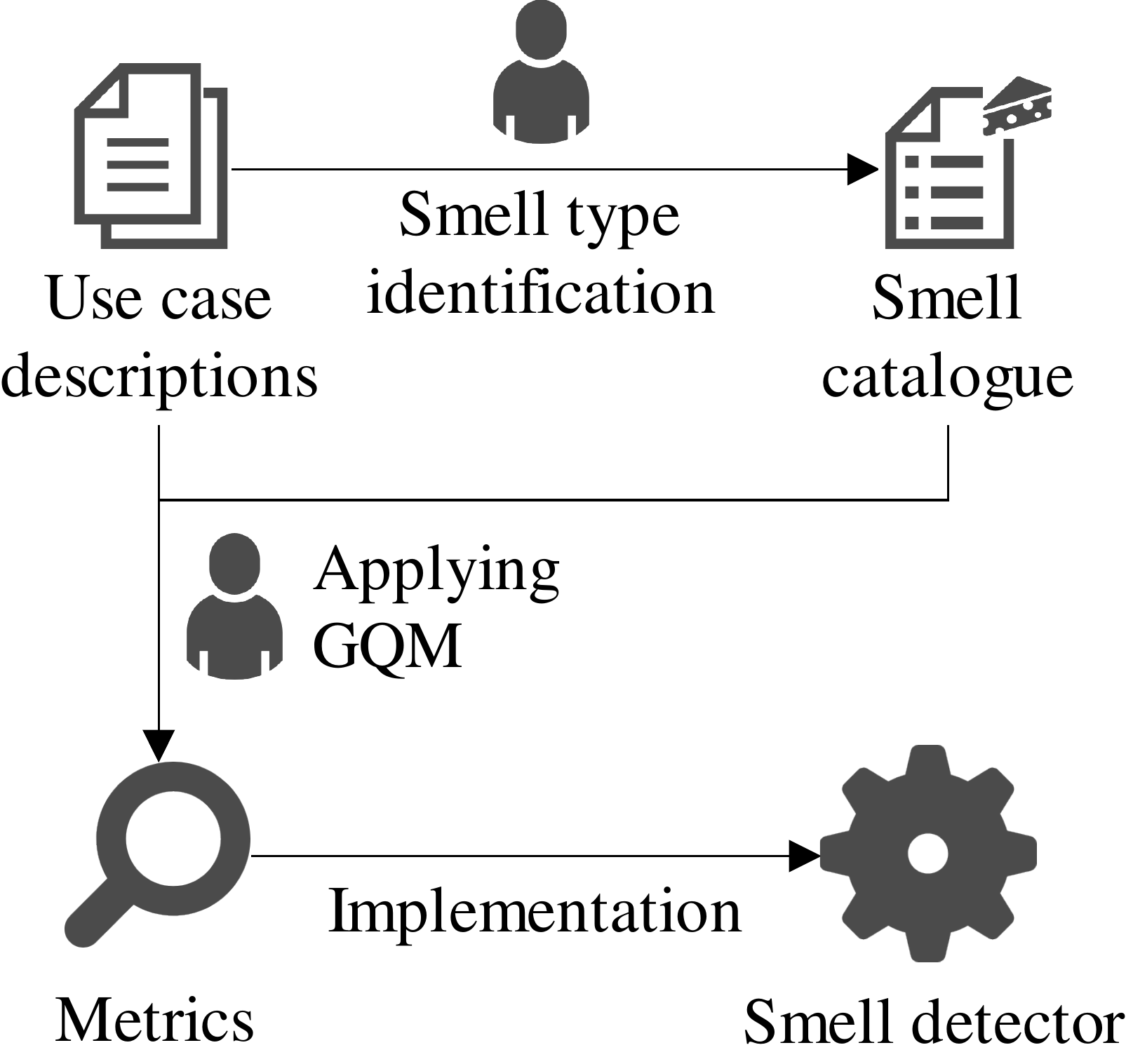}
  \caption{Overview of our approach.}\label{f:approach}
\end{figure}

In our approach, existing use case descriptions were analyzed, and bad smells in use case descriptions and metrics for automatically detecting them were derived.
The overview of our approach is shown in Fig.~\ref{f:approach}.
First, we collected examples of existing use case descriptions from existing resources such as via the Internet or textbooks.
Next, one of the authors identified the problems in the descriptions and abstracted them to extract bad smell types and compose a smell catalogue.
A catalogue of bad smells may be useful to watch bad smells with a bird's eye view, as well as to use as a checklist for human analysts.
In particular, we do not think that any bad smell can be automated to detect, and we can understand which bad smells can be automated to detect with the existing current technology that we can get easily.
These are reasons why we tried to make a catalogue at the first stage of this research.

Subsequently, we applied the GQM paradigm \cite{GQM} and derived metrics to identify the obtained smells.
The derived metrics were machine-understandable so that an automated smell detector was implemented.

Note that this work was done in Japanese.
We collected use case descriptions written in Japanese, and the smell identification and evaluation were conducted by native speakers of Japanese.
This decision benefited us to better identify smells from many use case descriptions.
All the example use case descriptions in this paper were originally written in Japanese and were directly translated into English.

The details of each study will be explained in the subsequent subsections.

\subsection{Collecting Examples}

\begin{table}[tb]\centering
  \caption{Samples of Use Case Descriptions}\label{t:usecases}
  \rowcolors{2}{lightgray!30}{}
  \begin{tabular}{lll} \toprule
    Name                                       & Domain & Source \\ \midrule
    Move piece on board                        & Game     & Search \\
    Print elterly relief call application form & Welfare  & Search \\
    Create new blog account                    & Web      & Textbook \\
    Search for product                         & Shopping & Image Search \\ \bottomrule
  \end{tabular}
\end{table}

Most samples of use case descriptions that we used for this study were retrieved using web searches with the query ``use case descriptions.''
We have used not only the normal Google search but Google Image search to find images representing a use case description.
We have also added a specific term of a domain such as ``welfare'' as an additional query to find domain-specific use case descriptions.
We continued the search until we collect enough samples that met the following two conditions:
  1) those written in natural language so that people can understand them, and
  2) those contents are separated by sections so that the location where smells are occurring can be easily pointed out.
In addition to the found samples via a web search, we also used the descriptions found in research papers and textbooks that we have already known.
Finally, we have gathered 30 sample use case descriptions.\footnote{The descriptions used for the experiment shown in Section~\ref{s:evaluation} were also collected here but kept unused for this study.}
These use case descriptions were chosen to cover a wide range of domains such as web application, health care, or welfare.
An excerpt list of samples are shown in Table~\ref{t:usecases}.

\begin{figure}[tb]
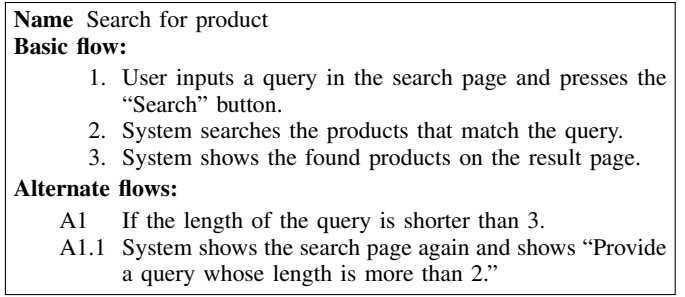
\centering
  {\small\fbox{\begin{minipage}[t]{0.98\linewidth}%
  \begin{LaTeXdescription}
    \item[Name] Search for product
    \item[Basic flow:]~\begin{enumerate}
      \item[1.] User inputs a query in the search page and presses the ``Search'' button.
      \item[2.] System searches the products that match the query.
      \item[3.] System shows the found products on the result page.
    \end{enumerate}
    \item[Alternate flows:]~\begin{enumerate}
        \item[A1\phantom{.1}] If the length of the query is shorter than 3.
        \item[A1.1] System shows the search page again and shows ``Provide a query whose length is more than 2.''
      \end{enumerate}
  \end{LaTeXdescription}
  \end{minipage}}}
  \caption{Example of a use case description.}\label{f:example}
\end{figure}
Figure~\ref{f:example} shows an example of the collected use case descriptions.
This use case description is for searching products, and its contents are separated into multiple sections: Name, Basic Flow, and Alternate Flows.

\subsection{Deriving Smells}

\newcommand{\SLIST}[2]{\item \textbf{#1} \begin{itemize}#2\end{itemize}}

\begin{table*}[tb]\centering
  \caption{Bad Smells in Use Case Descriptions}\label{t:smells}
  \begin{tabular}{c}\toprule\vspace{-1.5em}\\
    \begin{minipage}{.95\textwidth}\begin{multicols}{2}\begin{itemize}
        \SLIST{$C_{1,2}$: $\Pair{\Ambiguity}{\Section}$}{
          \item \Smell{Unordered Flow}\Impl
          \item \Smell{Origin-Free Exception Flow}\Impl
          \item \Smell{Origin-Free Alternate Flow}\Impl
        }
        \SLIST{$C_{1,4}$: $\Pair{\Ambiguity}{\Sentence}$}{
          \item \Smell{Unclear Feasibility}
          \item \Smell{Origin-Free Operation Result}
          \item \Smell{Sentence Interpretable as Multiple Meanings}
        }
        \SLIST{$C_{1,5}$: $\Pair{\Ambiguity}{\Word}$}{
          \item \Smell{Pronoun}\Impl
          \item \Smell{Omitted Word}
          \item \Smell{``Actor'' Actor}\Impl
          \item \Smell{Unexplained Main Actor}
          \item \Smell{Different Concepts by Same Word}
          \item \Smell{Omitted Attribute}
        }
        \SLIST{$C_{2,2}$: $\Pair{\Incorrectness}{\Section}$}{
          \item \Smell{Flow Does Not Meet Precondition}
          \item \Smell{Postcondition Not Satisfied}
          \item \Smell{Name Does Not Explain Content}
          \item \Smell{Under or Over Condition}
          \item \Smell{Much or Less Actors}
        }
        \SLIST{$C_{2,4}$: $\Pair{\Incorrectness}{\Sentence}$}{
          \item \Smell{Contradicted Sentences}
          \item \Smell{Behavior Ignores Condition}
        }
        \SLIST{$C_{3,1}$: $\Pair{\Granularity}{\Usecase}$}{
          \item \Smell{Multiple Situations}
        }
        \SLIST{$C_{3,2}$: $\Pair{\Granularity}{\Section}$}{
          \item \Smell{Multiple Exception Flows at an Exception Branch Condition}\Impl\Late\!\!\!\!\!\!\!\!\!\!\!\!\!\!\!\!\!\!\!\!
          \item \Smell{Multiple Alternate Flows at an Alternate Branch Condition}\Impl\Late\!\!\!\!\!\!\!\!\!\!\!\!\!\!\!\!\!\!\!\!
          \item \Smell{Multiple Roles of an Actor}
          \item \Smell{Multiple Actors of a Role}
        }
        \SLIST{$C_{3,4}$: $\Pair{\Granularity}{\Sentence}$}{
          \item \Smell{Long Sentence}\Impl
          \item \Smell{Short Sentence}\Impl
          \item \Smell{Sentence with Multiple Actions}\Impl
          \item \Smell{Relatively Over Qualified Sentence}\Impl
          \item \Smell{Relatively Under Qualified Sentence}\Impl
        }
        \SLIST{$C_{3,5}$: $\Pair{\Granularity}{\Word}$}{
          \item \Smell{Omitting Pre-Appeared Word}
          \item \Smell{Qualified Pre-Appeared Word}
        }
        \SLIST{$C_{4,3}$: $\Pair{\Redundancy}{\Flow}$}{
          \item \Smell{Multiple Flows with the Same Role}\Late
          \item \Smell{Flow Unrelated to Postcondition}\Late
          \item \Smell{Conditional Flow}\Late
        }
        \SLIST{$C_{4,4}$: $\Pair{\Redundancy}{\Sentence}$}{
          \item \Smell{Repeating the Same Noun}\Impl
        }
        \SLIST{$C_{4,5}$: $\Pair{\Redundancy}{\Word}$}{
          \item \Smell{Over-Qualified Word}
        }
        \SLIST{$C_{5,1}$: $\Pair{\Lack}{\Usecase}$}{
          \item \Smell{Non-Standalone Use Case}
        }
        \SLIST{$C_{5,2}$: $\Pair{\Lack}{\Section}$}{
          \item \Smell{Missing Actor Section}\Impl
          \item \Smell{Missing Exception Flows Section}\Impl
          \item \Smell{Missing Alternate Flows Section}\Impl
          \item \Smell{Missing Preconditions Section}\Impl
          \item \Smell{Missing Postconditions Section}\Impl
          \item \Smell{Missing Description Section}\Impl
          \item \Smell{Missing Name Section}\Impl
        }
        \SLIST{$C_{5,3}$: $\Pair{\Lack}{\Flow}$}{
          \item \Smell{Premature Exceptional Cases}
          \item \Smell{Premature Branch Condition}
        }
        \SLIST{$C_{5,4}$: $\Pair{\Lack}{\Sentence}$}{
          \item \Smell{Exception Flow without Return}\Impl
          \item \Smell{Unexplained Exception Flow}\Impl
          \item \Smell{Alternate Flow without Return}\Impl
          \item \Smell{Unexplained Alternate Flow}\Impl
          \item \Smell{Incomplete System Behavior}
          \item \Smell{Incomplete System Information}
        }
        \SLIST{$C_{5,5}$: $\Pair{\Lack}{\Word}$}{
          \item \Smell{Missing Action Target}
          \item \Smell{Missing Operation Procedure}
          \item \Smell{Unknown Origin}
        }
        \SLIST{$C_{6,2}$: $\Pair{\Misplacement}{\Section}$}{
          \item \Smell{Precondition in Basic Flow}
          \item \Smell{Postcondition in Basic Flow}
          \item \Smell{Exception Flow in Basic Flow}
          \item \Smell{Alternate Flow in Basic Flow}
        }
        \SLIST{$C_{7,5}$: $\Pair{\Inconsistency}{\Word}$}{
          \item \Smell{Synonym}\Late
        }
    \end{itemize}\end{multicols}\end{minipage}\\\vspace{-0.5em}~\\\bottomrule
  \end{tabular}%
\end{table*}

One of the authors read the collected use case descriptions carefully and identified their problems together with their reasons.
Then, smell types were derived according to the similarity of the reasons attached to the problems.
As a result, 248 problems were totally identified in the 30 use case descriptions, and 54 smell types were derived from them.
All the smell types are listed in Table~\ref{t:smells} according to the classification criteria, which will be explained later.
Note that Table~\ref{t:smells} lists up totally 60 smell types, and six smell types annotated with ``*'' in this table were not identified in this study and will be introduced later.

\begin{table}[tb]\centering
  \caption{Smell Characteristics}\label{t:characteristics}
  \rowcolors{2}{lightgray!30}{}
  \begin{tabular}{lp{5.8cm}} \toprule
    Characteristic & Description \\\midrule
    \Ambiguity     & The meaning cannot be determined uniquely, and its understanding requires effort. \\
    \Incorrectness & The content is incorrect or inconsistent. \\
    \Granularity   & The content is too rough or too detailed. \\
    \Redundancy    & There is an extra unnecessary part in the content. \\
    \Lack          & There is a missing necessary part. \\
    \Misplacement  & The content is located where it should not be located. \\
    {\Inconsistency}\Late & Two of the contents are inconsistent. \\\bottomrule
  \end{tabular}
\end{table}

\begin{table}[tb]\centering
  \caption{Smell Scope}\label{t:scope}
  \rowcolors{2}{lightgray!30}{}
  \begin{tabular}{lp{6.1cm}} \toprule
    Scope     & Description \\\midrule
    \Usecase  & A problem is found in an entire use case description. \\
    \Section  & A problem is found in a specific section such as Actors or Description. \\
    \Flow     & A problem is found in a specific flow or its subset representing a sequence of steps.
                A typical case is that two steps in it are inconsistent. \\
    \Sentence & A problem is found in a specific statement in a description. \\
    \Word     & A problem is found in a word in a description. \\\bottomrule
  \end{tabular}
\end{table}
We also found that the derived smell types could be categorized in two views.
The first view is the general reasons why the smells are problematic, which we call it \emph{smell characteristic} hereafter.
We identified six characteristics: \Ambiguity, \Incorrectness, \Granularity, \Redundancy, \Lack, and \Misplacement.
The meaning of each characteristic is explained in Table~\ref{t:characteristics}.
For example, 12 out of 54 smells were related to the ambiguity of some specific contents in a use case description, which were classified as \Ambiguity.
The second view is the \emph{scope} of smells.
We found that the scope of identified smell instances diversified in a wide variety from small-scoped ones such as word level to large-scoped ones such as an entire use case description.
Therefore, we also classified the smell types based on their scopes: \Usecase, \Section, \Flow, \Sentence, and \Word levels.
Their meanings are explained in Table~\ref{t:scope}.
Note that the smell characteristic of \Inconsistency, which is annotated with ``*'' in Table~\ref{t:characteristics}, was not identified in this study and will be introduced later.

\newcommand{\Level}[1]{\hspace{1em}$L_#1$\hspace{1em}}
\begin{table}[bt]\centering
  \caption{Smell Space}\label{t:space}
  {\tabcolsep=5pt\begin{tabular}{l|ccccc} \toprule
    \multicolumn{1}{r|}{Scope} & \Level{1} & \Level{2} & \Level{3} & \Level{4} & \Level{5} \\
    Characteristic & \Usecase & \Section & \Flow & \Sentence & \Word \\\midrule
    1. \Ambiguity     & $C_{1,1}$ & $C_{1,2}$ & $C_{1,3}$ & $C_{1,4}$ & $C_{1,5}$ \\
    2. \Incorrectness & $C_{2,1}$ & $C_{2,2}$ & $C_{2,3}$ & $C_{2,4}$ & $C_{2,5}$ \\
    3. \Granularity   & $C_{3,1}$ & $C_{3,2}$ & $C_{3,3}$ & $C_{3,4}$ & $C_{3,5}$ \\
    4. \Redundancy    & $C_{4,1}$ & $C_{4,2}$ & $C_{4,3}$ & $C_{4,4}$ & $C_{4,5}$ \\
    5. \Lack          & $C_{5,1}$ & $C_{5,2}$ & $C_{5,3}$ & $C_{5,4}$ & $C_{5,5}$ \\
    6. \Misplacement  & $C_{6,1}$ & $C_{6,2}$ & $C_{6,3}$ & $C_{6,4}$ & $C_{6,5}$ \\
    7. \Inconsistency & $C_{7,1}$ & $C_{7,2}$ & $C_{7,3}$ & $C_{7,4}$ & $C_{7,5}$ \\\bottomrule
  \end{tabular}}
\end{table}
As mentioned before, we have two orthogonal views: Characteristic and Scope to categorize bad smells.
Considering a view as a coordinate axis, we can have a two-dimensional space called \emph{smell space}.
As shown in Table~\ref{t:space}, we may write a category of bad smells with a two dimensional coordinate or vector like $C_{3,4} = \Pair{\Granularity}{\Sentence}$ where the values of Characteristic-coordinate = \Granularity and Scope-coordinate = \Sentence.

\begin{figure}[tb]\centering
  {\small\fbox{\begin{minipage}[t]{8.7cm}%
  \begin{LaTeXdescription}
    \item[Name:] Long Sentence
    \item[Characteristic:] \Granularity
    \item[Scope:] \Sentence
    \item[Symptom:]
      A sentence is relatively long to the other ones in a section.
      The target sentence may contain too much information or may contain unnecessary information, which decreases the understandability of the sentence.
      It should be separated so as to be easier to understand.
    \item[How to Detect:]
      Checks whether the number of characters in a sentence (length of a sentence) exceeds a threshold which is calculated by the distribution of the lengths of the sentences among the target use case description
  \end{LaTeXdescription}
  \end{minipage}}}
  \caption{Smell description of \Smell{Long Sentence}.}\label{f:smell-example}
\end{figure}
Figure~\ref{f:smell-example} shows an example of the documentation of our bad smell catalogue.
It contains the following sections:
\begin{LaTeXdescription}
  \item[Name:]
    The name of the bad smell.
    It should be a sufficiently descriptive and unique name.
  \item[Characteristic and Scope:]
    The characteristic and the scope of the bad smell, which are one of the items listed in Table~\ref{t:characteristics} and one in Table~\ref{t:scope}, respectively.
    These also specify the category of the smell, which is based on Table~\ref{t:space}.
    The smell example shown in Fig.~\ref{f:smell-example} belongs to the category of $\Pair{\Granularity}{\Sentence}$.
  \item[Symptom:]
    Explanation on the smell, including surface features of the descriptions and the problems that the smell can cause.
    It provides intuitive reasons why these surface features could be problematic.
  \item[How to Detect:]
    More concrete descriptions on the surface features of the descriptions to detect the bad smell by manual and/or machine.
    This section can be used as a checklist like \cite{phalp-sqj2007}, \cite{torner-isese2006}, etc.
    Some of the features can be automated to detect and we have implemented them as the smell detector.
\end{LaTeXdescription}

\begin{table}[bt]\centering
  \caption{Numbers of Identified Smell Instances}\label{t:instances}
  {\tabcolsep=5pt\begin{tabular}{l|ccccc|c} \toprule
    Characteristic & \Usecase & \Section & \Flow & \Sentence & \Word & Total \\\midrule
    \Ambiguity     &          &    23    &       &   12      &   25  & 60 \\
    \Incorrectness &          &    21    &       &    4      &       & 25 \\
    \Granularity   &   3      &     2    &       &   30      &    3  & 38 \\
    \Redundancy    &          &          &       &    3      &    1  & 4 \\
    \Lack          &   1      &    50    &  19   &   25      &   13  & 108 \\
    \Misplacement  &          &    13    &       &           &       & 13  \\\midrule
    Total          &   4      &   109    &  19   &   74      &   42  & 248 \\\bottomrule
  \end{tabular}}
\end{table}
The identified smell instances (problems) were distributed over different smell categories, i.e., the smell space.
Table~\ref{t:instances} shows how many instances were classified in the categories defined in the smell space.
Although the smell instances in \Section and those for \Lack were identified at most, those for other scopes and characteristics were also identified.

\subsection{Deriving Metrics via GQM}

Next, to implement an automated smell detector, metrics of a use case description are derived by applying Goal-Question-Metric~(GQM) paradigm \cite{GQM} to the identified smells mentioned in the last subsection.
More specifically, GQM application was conducted according to the following steps:
\begin{enumerate}
  \item Top two layers were used as \emph{Goal} part, and the goals in these layers were systematically derived.
         The categories in the smell space, i.e., pairs of the scope and the characteristic of smells, were used for the root goals.
         The specific types of the smells were used for the sub goals of these root goals in the second layer.
  \item As \emph{Question} part, questions were derived as to which features should be examined to detect the instances of each smell in the sub goal layer.
  \item Finally, for each question, metrics to be used to answer the questions were derived as leaves in \emph{Metric} layer.
\end{enumerate}

\begin{figure}[tb]\centering
  \includegraphics[width=\linewidth]{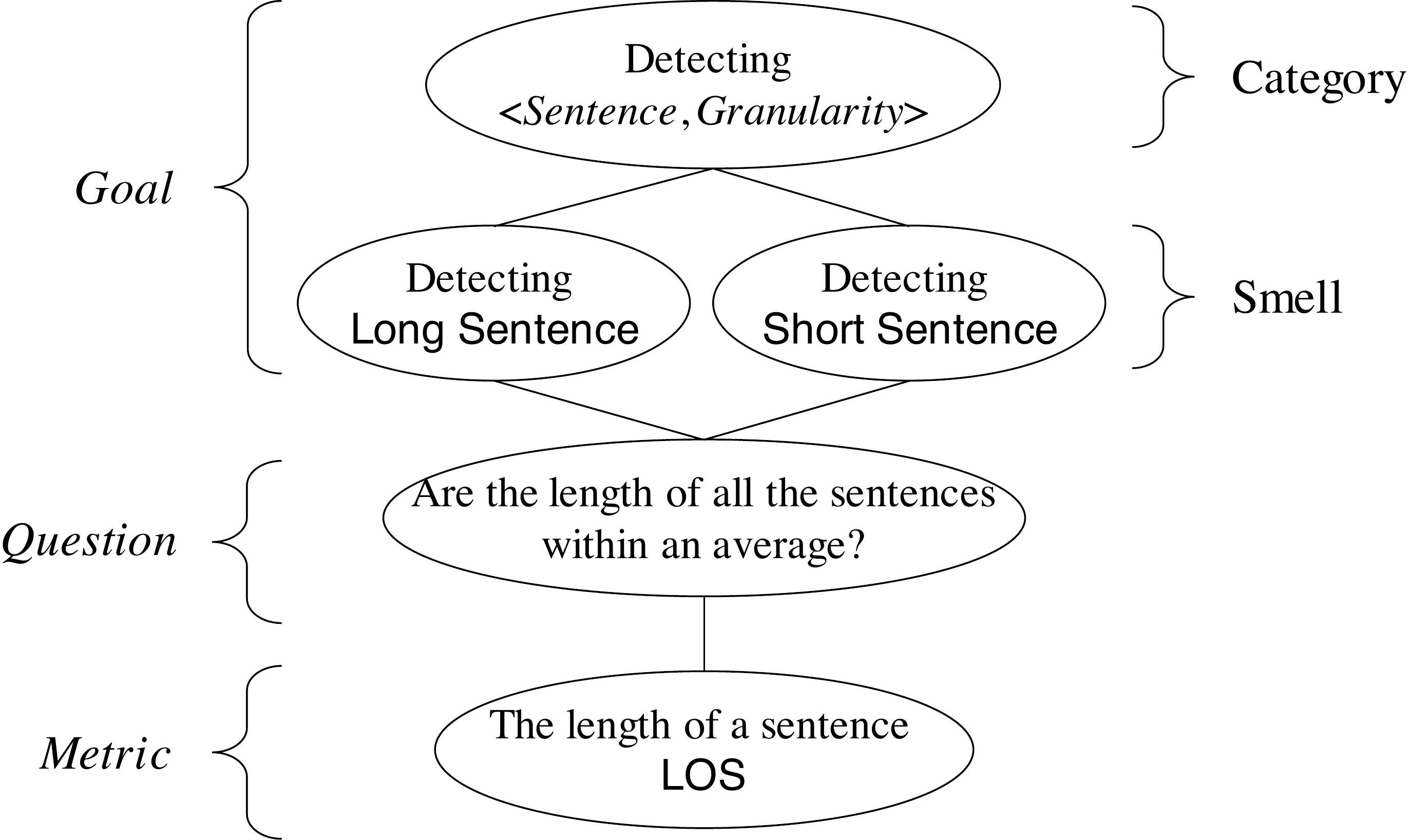}
  \caption{Excerpt example of GQM application for \Smell{Long Sentence} and \Smell{Short Sentence}.}\label{f:gqm}
\end{figure}

\begin{table}[bt]\centering
  \caption{Numeric Metrics to Detect Smells}\label{t:metrics}
  \rowcolors{2}{lightgray!30}{}
  \begin{tabular}{ll} \toprule
    Name                        & Description \\\midrule
    \Metric{NOP}                & The \U{n}umber \U{o}f \U{p}ronons \\
    \Metric{NOW}(\textit{word}) & The \U{n}umber \U{o}f the \U{w}ord \textit{word} \\
    \Metric{NOEFR}\Late         & The \U{n}umber \U{o}f \U{e}xception \U{f}lows with the same \U{r}eason \\
    \Metric{NOAFR}\Late         & The \U{n}umber \U{o}f \U{a}lternate \U{f}lows with the same \U{r}eason \\
    \Metric{LOS}                & The \U{l}ength \U{o}f a \U{s}entence \\
    \Metric{NOV}                & The \U{n}umber \U{o}f \U{v}erbs \\
    \Metric{NOM}                & The \U{n}umber \U{o}f \U{m}odifiers \\
    \Metric{NON}(\textit{noun}) & The \U{n}umber \U{o}f the \U{n}oun word \textit{noun} \\\bottomrule
  \end{tabular}
\end{table}

\begin{table}[bt]\centering
  \caption{Predicates to Detect Bad Smells}\label{t:predicates}
  \rowcolors{2}{lightgray!30}{}
  {\tabcolsep=1.5pt\begin{tabular}{lp{4.7cm}} \toprule
    Name                                       & Description \\\midrule
    \Predicate{BasicFlowNumbered?}             & Are all the steps in the given flow numbered? \\
    \Predicate{ExceptionFlowsNumbered?}        & Are all the steps in the exception flows numbered? \\
    \Predicate{AlternateFlowsNumbered?}        & Are all the steps in the alternate flows numbered? \\
    \Predicate{BasicFlowOrdered?}              & Are all the steps in the basic flow well ordered? \\
    \Predicate{ExceptionFlowsOrdered?}         & Are all the steps in the exception flows well ordered? \\
    \Predicate{AlternateFlowsOrdered?}         & Are all the steps in the alternate flows well ordered? \\
    \Predicate{BasicFlowStartWith1?}           & Does the basic flow start with 1? \\
    \Predicate{ExceptionFlowsStartWith1?}      & Is the index of the first step of the exception flows 1? \\
    \Predicate{AlternateFlowsStartWith1?}      & Is the index of the first step of the alternate flows 1? \\
    \Predicate{ExceptionFlowsOriginDescribed?} & Is the origin of the exception flows described? \\
    \Predicate{AlternateFlowsOriginDescribed?} & Is the origin of the alternate flows described? \\
    \Predicate{ActorSectionExist?}             & Does the Actors section exist? \\
    \Predicate{ExceptionFlowsSectionExist?}    & Does the Exception Flows section exist? \\
    \Predicate{AlternateFlowsSectionExist?}    & Does the Alternate Flows section exist? \\
    \Predicate{PreconditionsSectionExist?}     & Does the Precondition section exist? \\
    \Predicate{PostconditionsSectionExist?}    & Does the Postcondition section exist? \\
    \Predicate{OverviewSectionExist?}          & Does the Overview section exist? \\
    \Predicate{NameSectionExist?}              & Does the Name section exist? \\
    \Predicate{ExceptionFlowsReturnExist?}     & Is it described where the exception flows return to? \\
    \Predicate{AlternateFlowsReturnExist?}     & Is it described where the alternate flows return to? \\
    \Predicate{ExceptionFlowsReasonExist?}     & Are the conditions when exception flows are executed described? \\
    \Predicate{AlternateFlowsReasonExist?}     & Are the conditions when alternate flows are executed described? \\\bottomrule
  \end{tabular}}
\end{table}

An excerpt example of the derivation of metrics by applying GQM is shown in Fig.~\ref{f:gqm}.
\begin {enumerate}
  \item As a root goal, we considered the category of $\Pair{\Granularity}{\Sentence}$ and derived the goal ``Detecting smells of an inappropriate granularity in a sentence'' as a root goal.
  Then, we considered \Smell{Long Sentence} and \Smell{Short Sentence} as smells related to an inappropriate granularity in a sentence, and derived two sub goals ``Detecting \Smell{Long Sentence}'' and ``Detecting \Smell{Short Sentence}'' at the second layer.
  \item In order to detect \Smell{Long Sentence}, the main question is whether the length of a given sentence is regarded as within the average, and the question ``Are the length of all the sentences within an average?'' was derived.
  \item To answer the question above, we need to measure the length, i.e., the number of characters, of a sentence to determine whether the sentence length is within an average range of the sentence lengths.
\end{enumerate}
Metrics derived as leaves could be implemented as \Metric{LOS}~(The \U{l}ength \U{o}f a \U{s}entence).
The same approach was applied to the other smells to derive metrics.
There are two types of metrics to be derived: not only those that output a numeric value such as \Metric{LOS} as explained above, but also those that determine whether the given content of a use case description satisfies a specific condition.
We call the formers and the latters \emph{numeric metrics} and \emph{predicates}, respectively.
Predicates can be implemented as metrics that output a boolean value.

By applying GQM, machine-measurable metrics were derived for 22 smell types.
In deriving machine-measurable metrics, we limited the analysis scope within the syntactic analysis of natural language descriptions and structural analyses of a use case description.
Of the smells shown in Table~\ref{t:smells}, those annotated with ``$\diamond$'' are those that are detectable using the derived smells.
Tables~\ref{t:metrics} and \ref{t:predicates} list the derived numeric metrics and predicates.
Note that, similarly to Table~\ref{t:smells}, numeric metrics annotated with ``*'' shown in Table~\ref{t:metrics} were not derived from the analysis in this section, and they were derived during the experiment.
The details will be explained in the next section.

Instead of these 22 smell types, we failed to derive machine-measurable metrics for the other 32 smell types.
A typical reason why machine-measurable metrics derivation was failed for such smell types is that they were related to the semantic concepts and/or domain knowledge.
For example, consider trying to derive metrics for \Smell{Multiple Actors in a Role}, which indicates that two different actors actually have the same role.
For example, an actor named ``Family'' in the context of medical care may have the role of liaison between a patient and a doctor, which causes a situation that two actors having different names have the same semantic role.
However, to detect smells of this type, we need to extract the semantic role of actors, which is unable only with syntactic and structural analyses.

To evaluate the metrics that could be automated, we have implemented a prototype of the automated smell detector with object-oriented script language Ruby\footnote{\url{https://www.ruby-lang.org/}}.
The reasons why we used the script language are that
\begin{enumerate}
  \item we can add new metrics and change the metrics easily and flexibly, and
  \item we can integrate the functions of the smell detector to the other use case supporting tools.
\end{enumerate}
For implementing the detector, we used Mecab\footnote{\url{https://taku910.github.io/mecab/}} as a morphonological analyzer and NEologd\footnote{\url{https://github.com/neologd/mecab-ipadic-neologd}} \cite{NEologd} as a dictionary to analyze Japanese sentences in use case descriptions.
The outputs of the detector are the information on detected bad smells including locations, used metrics, etc.
The detection rules of the smell types annotated with ``$\diamond$'' shown in the list in Table~\ref{t:smells} have been implemented in the detector.

\begin{figure}[tb]\centering
  \includegraphics[width=8cm]{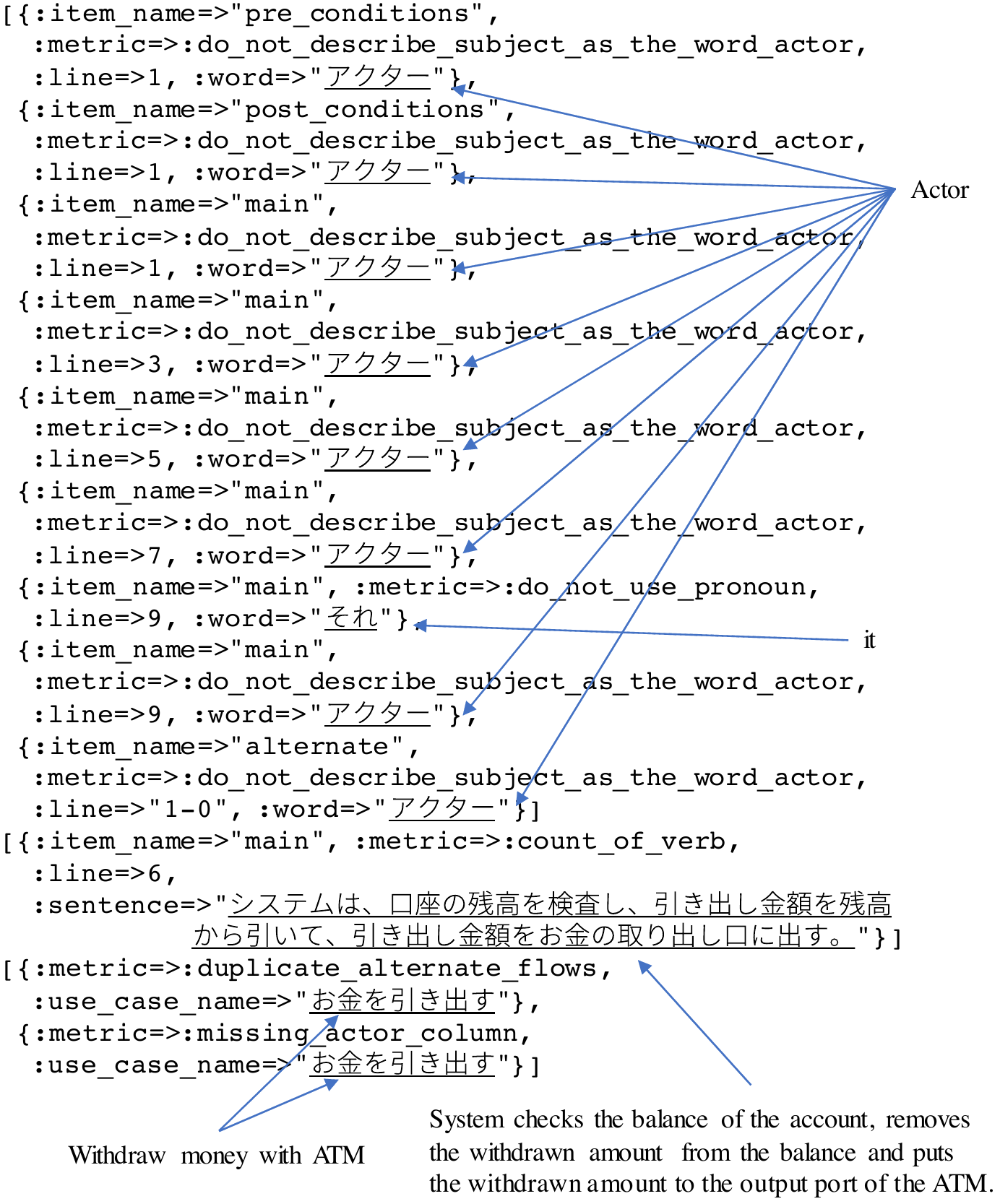}
  \caption{Snapshot of the tool output.}\label{f:tool-output}
\end{figure}
\begin{figure*}[tb]\centering
  \includegraphics[width=15cm]{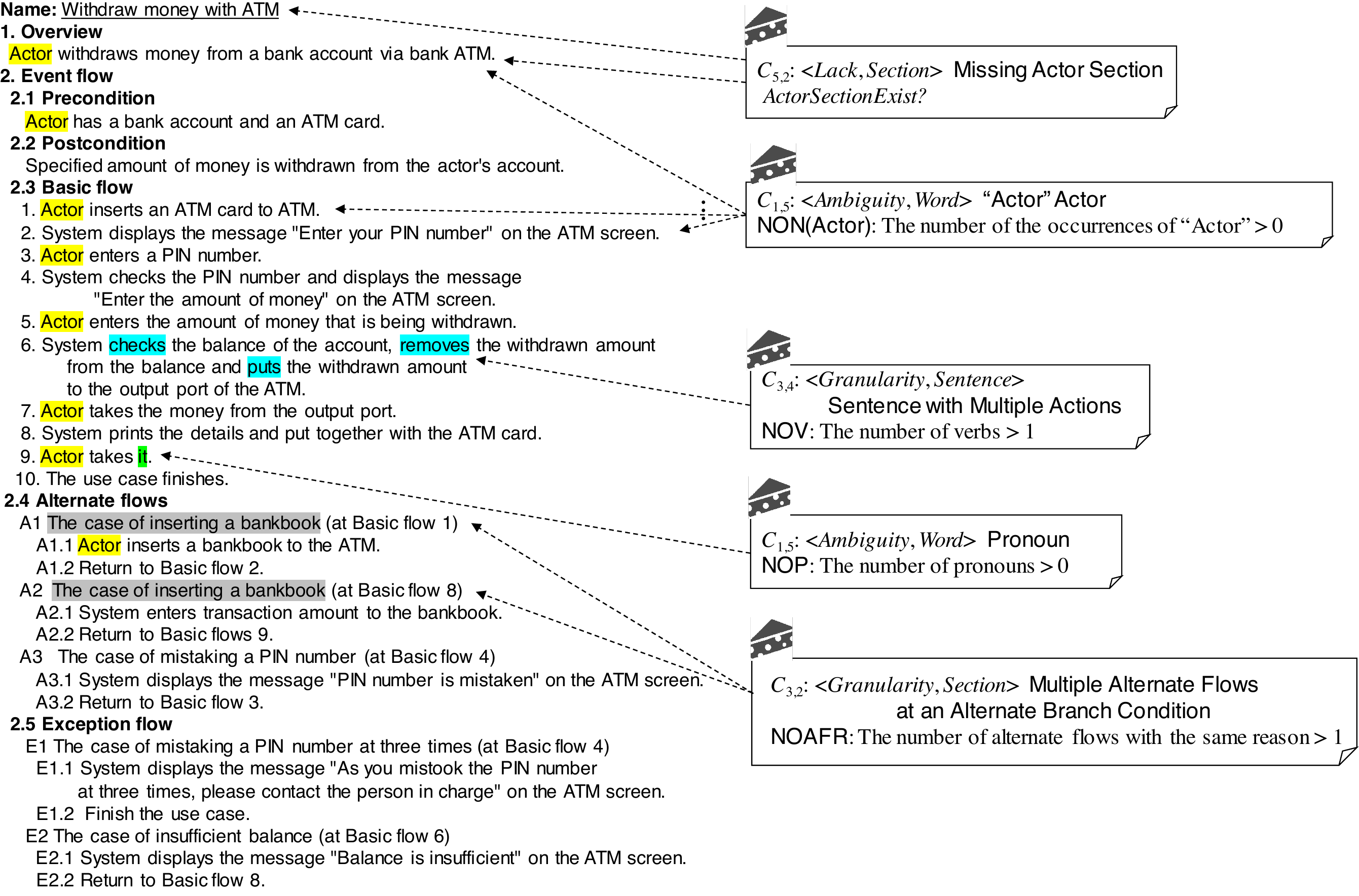}
  \caption{Example result of the smell detection.}\label{f:detection-example}
\end{figure*}

Figure~\ref{f:tool-output} shows the snapshot of the output of our detector for the use case ``Withdraw money with ATM'', which is shown in the left part of Fig.~\ref{f:detection-example}.
The output follows the JSON data format\footnote{https://www.json.org/}.
A string parenthesized with ``\texttt{\{}'' and ``\texttt{\}}'' denotes an occurrence of detected bad smells, and it normally consists of four lines.
The first line headed with ``\texttt{item\_name}'', the second with ``\texttt{metric}'', the third with ``\texttt{line}'', and the fourth with ``\texttt{word}'', ``\texttt{sentence}'', or ``\texttt{flow}'' represent the location of the occurrence in the use case description, the used metrics to detect it, the line number of it in the location, and its string data, respectively.
In the example of the first occurrence of the detected bad smell, it appeared in the section ``Precondition'', the metrics \Metric{NON}(``actor'') (more comprehensively shown as ``do not describe subject as the word actor'') was used, it appeared in the first line of the precondition section and the word ``Actor'' in the line was caused the bad smell.
Since our metrics work for Japanese, our tool picked up a Japanese word, a Japanese sentence, and a set of Japanese sentences (a flow) as a bad smell.
In the right side of Fig.~\ref{f:tool-output}, the readers can find English translation of the detected underlined Japanese words and sentences.

Figure~\ref{f:detection-example} shows the example of a use case ``Withdraw money with ATM'' and its detection result.
The detected bad smells are listed in the right side of the figure.
The first bad smell is missing actor sections belonging to the category $C_{5,2}$, whose characteristic and scope are \Lack and \Section, respectively.
To detect this bad smell, the predicate \Predicate{ActorSectionExist?} in Table~\ref{t:predicates} returned the value False because our detector found that there were no sections on the specification of actors in the use case description.
The category $C_{1,5}$ of bad smells appeared in the eight sentences, whose subject is the word ``actor''.
If this use case was related to multiple actors, the word ``actor'' was ambiguous because we could not decide which actors the word ``actor'' denoted.
Our detector decided that the eight occurrences of the word ``actor'' caused a bad smell \Smell{``Actor'' Actor} using the metrics \Metric{NON}(``actor''), i.e., the number of the occurrences of ``actor'', in Table~\ref{t:metrics}.
The sixth sentence in ``2.3 Basic Flows'' section was decided to have the bad smell \Smell{Sentence with Multiple Actions} because it included three verbs ``checks'', ``removes'', and ``puts'' and thus was a compound sentence.
The usage of pronouns caused ambiguity, and our detector detected the pronoun ``it'' in the ninth sentence of Basic flows.
The last category of the detected bad smell was \Smell{Multiple Alternate Flows at an Alternate Branch Condition}.
In ``2.4 Alternate flows'' section, the branch conditions to alternate actions of A1 and A2 were the same.
It means that A1 and A2 should be executed under an alternate condition, and for easiness to read, they should be modularized together under the condition.

\section{Analytic and Experimental Evaluation}\label{s:evaluation}
In this section, we discuss the evaluation of our approach: the developed catalogue and the automated smell detector.

\subsection{Research Questions}\label{ss:rq}

First of all, we set up the following research questions to validate our approach.

\begin{LaTeXdescription}
  \item[\RQ{1}:] Can our catalogue cover various bad smells?
  \item[\RQ{2}:] Can our automated detector indicate all of the occurrences of bad smells correctly?
\end{LaTeXdescription}

\RQ{1} is concerned with the coverage of our catalogue.
To validate it from two views: an analytic view and an experimental one.
In the former one, we collect the existing popular checklists of use case descriptions and compared ours with them.
As for the latter one, we ask the study participants to find bad smells from some examples of use case descriptions and assess if their results would be included in our catalogue or not.
Thus, \RQ{1} can be refined into the following two:
\begin{LaTeXdescription}
\item[\RQ{1--1}:] Have our catalogue included the bad smells that study participants have found in use case description examples?
\item[\RQ{1--2}:] Can our catalogue cover some existing popular checklists?
\end{LaTeXdescription}

To respond \RQ{2}, we adopt an experimental comparative approach where we compare the correct set of bad smells with the results that our detector indicates.
We reuse as a correct set the results that the study participants produced in \RQ{1}.

\begin{figure}[tb]\centering%
  \includegraphics[width=0.9\linewidth]{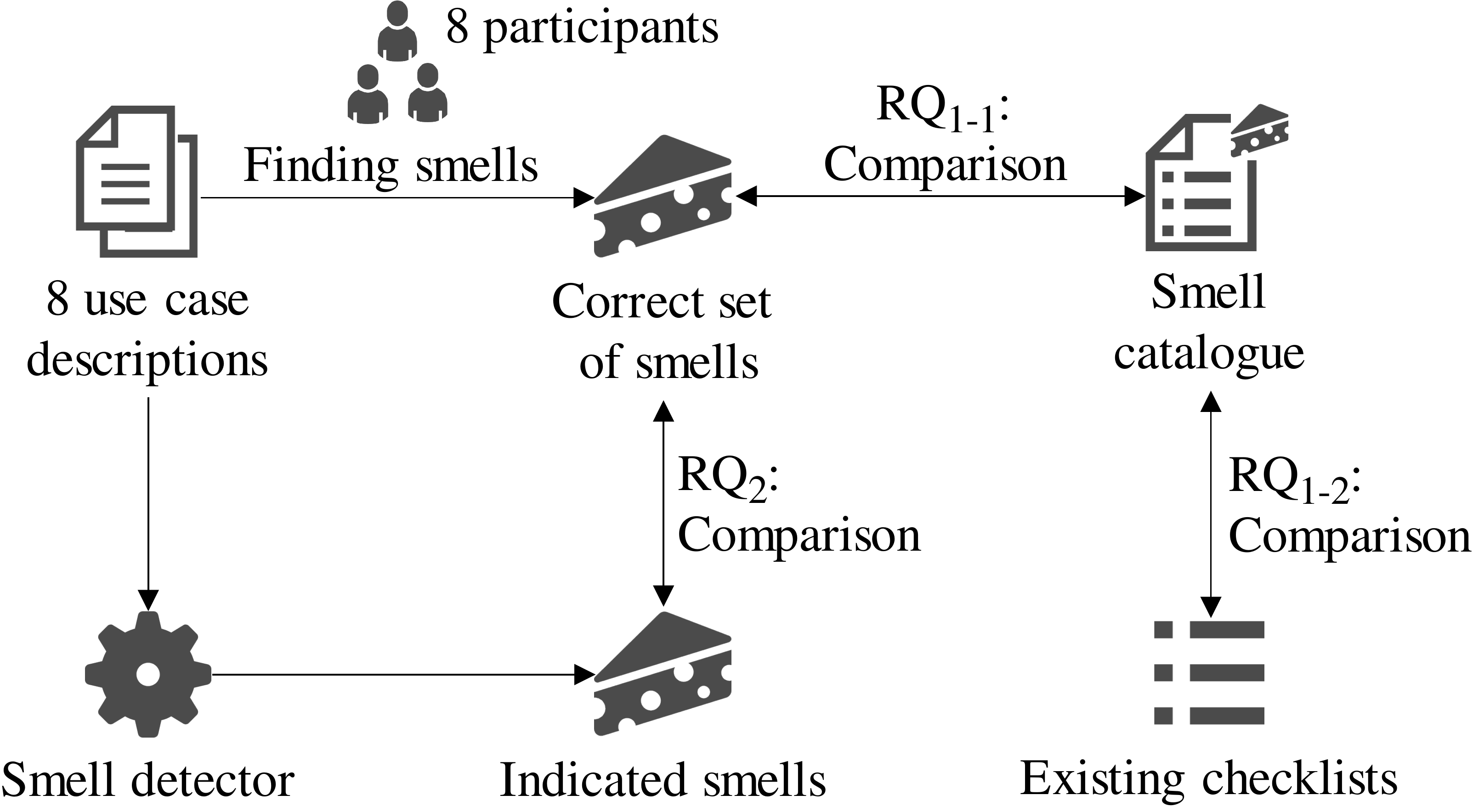}
  \caption{Overview of the evaluation process.}\label{f:eval-overview}
\end{figure}

The overview of our evaluation process is shown in Fig.~\ref{f:eval-overview} and its details will be mentioned in the next subsection.

\subsection{Evaluation Procedure}

As shown in Fig.~\ref{f:eval-overview}, we had eight examples of use case descriptions, which were different from the previous 30 examples that we used to develop the catalogue and the metrics, and eight study participants which had experienced in use case modeling.
The study participants consisted of two undergraduate and five graduate students, and one faculty in the department of computer science majoring in software quality and/or requirements engineering, who were gathered via the authors' network.
They have half to 15 years of experiences in use case modeling.
Our participants were asked to find in the examples the poor descriptions that might cause any of various problems in some aspects, e.g., understandability, realizability, implementability, consistency, etc.
They were also required to clarify the reasons why they judged their findings as poor descriptions.
We first gathered all the participants at the same time in a single room and conducted a face-to-face meeting.
During the meeting, we explained the overview of the experiment, distributed printed sheets of use case descriptions, provided 30 min to find the examples, and responded questions from the participants.
After the meeting, participants were allowed to complete to find examples separately without any time limitation.
The examples were specified in the printed sheets as their location and reason.
We analyzed the reasons for the found poor descriptions and tried to classify them into bad smells of our catalogue.
We regarded the classification results as a correct set of bad smells included in the examples.
The correct set is used for responding \RQ{1--1} and \RQ{2}.

These eight examples have been collected from the Internet, and they were in the various problem domains.
It is difficult to create a complete and real correct set of bad smells because the judgment of bad smells or not is subjective to human analysts.
Thus, we took the poor descriptions that at least one of the eight study participants considered as poor descriptions so that we could capture the candidates of bad smell instances as many as possible.
One of the authors confirmed all the obtained smell instances.
During the confirmation, only one of them were excluded because that was based on a misreading of the description.

As for \RQ{1--1}, some of the bad smells in the correct set could not be classified into our catalogue, and we consider these bad smells as the new bad smells that should be added to the catalogue.
We counted the number of the new bad smells to respond \RQ{1--1}.

\begin{table}[bt]\centering
  \caption{Comparison with Existing Checklists}\label{t:catalogue-comparison}
  \begin{tabular}{lrrrr}\toprule
              &                & \# smell types not included \\
    Checklist & \# smell types & in our catalogue \\\midrule
    Phalp \etal \cite{phalp-sqj2007}         & 13 & $2 \To 1$ \\
    T\"{o}rner \etal \cite{torner-isese2006} & 12 & $2 \To 1$ \\
    Anda and Sj{\o}berg \cite{anda-seke2002} & 27 & $6 \To 4$ \\\midrule
    & & \# smell types not included \\
    & & in \cite{phalp-sqj2007}, \cite{torner-isese2006} or \cite{anda-seke2002} \\\midrule
    Our catalogue & $54 \To 60$ & $4 \To 6$ \\\bottomrule
  \end{tabular}
\end{table}

\begin{table*}[bt]\centering
  \caption{Precision and Recall Values for Bad Smell Detection}
  \label{t:accuracy}
  \begin{tabular}{lrrrr}\toprule
    Category & Precision & Recall & \# detector indicated & \# smell instances in the correct set \\ \midrule
    $\Pair{\Ambiguity}{\Section}$    & 0.833 & 1.00 & 12 & 10 \\
    $\Pair{\Ambiguity}{\Word}$       & 0.522 & 1.00 & 23 & 12 \\
    $\Pair{\Granularity}{\Section}$  & $\text{N/A} \To 1.00$ & $0 \To 1.00$ & $0 \To 1$ & 1 \\
    $\Pair{\Granularity}{\Sentence}$ & 0.385 & 1.00 & 26 & 10 \\
    $\Pair{\Redundancy}{\Sentence}$  &   N/A & N/A  &  0 &  0 \\
    $\Pair{\Lack}{\Section}$         & 0.696 & 1.00 & 23 & 16 \\
    $\Pair{\Lack}{\Sentence}$        &  1.00 & 1.00 &  4 & 4 \\ \midrule
    Total & $0.591 \To 0.596$ & $0.981 \To 1.00$ & $88 \To 89$ & 53 \\ \bottomrule
    \end{tabular}
\end{table*}

To respond \RQ{1--2}, we selected three popular checklists made by
 1) Phalp \etal \cite{phalp-sqj2007},
 2) T\"{o}rner \etal \cite{torner-isese2006}, and
 3) Anda and Sj{\o}berg \cite{anda-seke2002},
and tried to compose a mapping between our catalogue and their check items.
The mapping allows us to analyze if the bad smells of our catalogue cover their check items or not.
More precisely, we focus on the bad smells of our catalogue that cannot be mapped to any of their check items (we say, a bad smell is not included in a checklist) and on the check items that cannot be mapped to any of our bad smells (a checklist item is not included in our catalogue).
Since the granularity of our catalogue may be different from that of their check items, the mapping cannot necessarily be one-to-one, but one-to-many or many-to-many.

We applied our smell detector to the eight examples and compared the results with the correct set.
More concretely, we listed up
\begin{enumerate}
  \item the bad smell instances that were included in the correct set and also indicated by the detector (true positives: $\TP$),
  \item the bad smell instances included in the correct set but not indicated by the detector (false negatives: $\FN$), and
  \item the descriptions not included in the correct set but indicated by the detector (false positives: $\FP$).
\end{enumerate}

To respond \RQ{2}, we calculate a precision ration and recall one as follows:
\begin{align*}
  \mathit{Precision} &= \frac{|\TP|}{|\TP| + |\FP|}, &
  \mathit{Recall} &= \frac{|\TP|}{|\TP| + |\FN|}.
\end{align*}

\subsection{Results}

\subsubsection{Comparison with the Correct Set (Response to \RQ{1--1})}

The experiment that our eight study participants conducted made us new six bad smells that were not included in the first version of our catalogue, the bad smells annotated with ``*'' shown in Table~\ref{t:smells}.
As a result, we have two versions of bad smell catalogue; the first version has 54 bad smells, and the improved version does 60 bad smells, and furthermore, we have developed the metrics to indicate these six newly added bad smells in the same way, i.e., using GQM approach.
As a result, we could finally get 30 metrics, which respectively consist of 8 numeric metrics and 22 predicates shown in Tables~\ref{t:metrics} and \ref{t:predicates}, for the improved version of our catalogue, by adding new two numeric metrics, which are annotated with ``*'' in Table~\ref{t:metrics}, to detect two of the newly added bad smells.
In the later subsections of responding \RQ{1--2} and \RQ{2}, we compare these two versions of catalogue with the existing checklists~(\RQ{1--2}).
We also use two versions of metrics to respond \RQ{2}.

\subsubsection{Comparison with Existing Checklists (Response to \RQ{1--2})}\label{sss:rq1-2}

Table~\ref{t:catalogue-comparison} shows the comparative results with the existing three checklists.
The checklists of Phalp \etal \cite{phalp-sqj2007}, T\"{o}rner \etal \cite{torner-isese2006}, and Anda and Sj{\o}berg \cite{anda-seke2002} have 13, 12, and 27 bad smell types, respectively.
Some cells in the table include two numerals together with the symbol ${\To}$.
The numeral in the left hand side of ${\To}$ stands for the value for the first version of our catalogue, and the right does for its improved version.
For example, Phalp \etal \cite{phalp-sqj2007} has two bad smell types that are not included in the first version of our catalogue, and they decrease to one for the improved version.

On the contrary, at first our catalogue had four bad smells that are not included in any of these checklists, and it could be improved so that it has six bad smell types not included in any of these existing checklists:
  \Smell{Omitted Attribute},
  \Smell{Multiple Exception Flows at an Exceptional Branch Condition},
  \Smell{Multiple Alternate Flows at an Alternate Branch Condition},
  \Smell{Multiple Roles of an Actor},
  \Smell{Multiple Actors of a Role}, and
  \Smell{Non-Standalone Use Case}.

On the other hand, we do not have one bad smell \Smell{Use Case Decomposition} (C7), where several activity flows should be extracted and moved into a new use case, in the checklist of T\"{o}rner \etal, and one bad smell \Smell{Consideration of Alternatives: Viable}, where alternate flows cannot be really executed, in Phalp \textit{et al.}
Four bad smells
  1) \textsf{Incorrect description of actors or wrong connection between actor and use case},
  2) \textsf{Inconsistencies between diagram and descriptions, inconsistent terminology, inconsistencies between use cases, or different level of granularity},
  3) \textsf{Actors that do not derive value from/provide value to the system}, and
  4) \textsf{Use cases with functionality outside the scope of the system or use cases that duplicate functionality}
of Anda and Sj{\o}berg are not also included in the catalogue.

\subsubsection{Comparison the Results of Tool Application with the Correct Set (Response to \RQ{2})}

Table~\ref{t:accuracy} shows the results for each bad smell category included in the eight examples.
It also includes the number of indication by the smell detector (\# detector indicated) and the number of the bad smell instances actually included in the examples (\# smell instances in the correct set), which our study participants indicated.
Totally, we found that the examples actually included 53 bad smell instances, and our tool could indicate 88 bad smell instances.
52 of the 88 bad smell instances were in the correct set, i.e., $\TP$.
For example for the bad smell category $\Pair{\Ambiguity}{\Section}$, we got the 12 bad smell instances that the detector indicated, and they included all of the ten correct bad smell instances.

Some of the cells in the table include two numerals with the symbol ${\To}$.
As mentioned in Section~\ref{ss:rq}, through constructing the correct set we found new additional six bad smell types and added them to the catalogue, and accordingly, we added several metrics so as to indicate these new bad smells.
The numerals in the left hand side of ${\To}$ stand for the results obtained by the detector before adding the new metrics and the right represent the result by the detector improved by adding the metrics.
In total, our improved detector indicated 89 bad smell instances, 53 of the 89 were correct.

High recall values could be partially satisfied to \RQ{2}, while lower precision values could not be sufficient.
We found two major reasons for the lower precision values.
The first one was the detection of ambiguous words: $\Pair{\Ambiguity}{\Word}$.
We consider the word ``actor'' is ambiguous because some use cases had multiple actors.
However, in the case of the use cases that have only one actor, our study participants did not consider that the word ``actor'' appearing in the sentences was ambiguous because they could identify correctly an object that the word ``actor'' denotes.
The second reason was related to the bad smell category $\Pair{\Granularity}{\Sentence}$.
We used the relative length of sentences as one of the metrics to detect smells related to the category $\Pair{\Granularity}{\Sentence}$.
Some sentences included illustrative phrases, and as a result, they came to be relatively long to the other sentences.
However, our study participants judged that these sentences including illustrative phrases were easier to understand, even if they were long.
This is the reason why our detector indicated bad smells incorrectly.

\subsection{Threats to Validity}\label{ss:threats}

\subsubsection{External Validity}
External validity is related to the generality of the obtained conclusions.
In this experiment, although we used only eight use case descriptions, their problem domains were different, and we believe that they covered varieties of the problem domain in a certain degree.

To evaluate the quality of our produced catalogue, we compared it with the other existing catalogues.
However, to compose the catalogue, we used a limited set of use case descriptions ($30 + 8$).
In fact, during the evaluation process, we used different eight use case descriptions, which our study participants analyzed, and it led to adding new six smells.
It means that the entries of our catalogue can be newly found and added whenever we focus on new use case descriptions.
We do not think that our catalogue would be complete or be a saturated catalogue.
However, our two orthogonal views, i.e., Characteristic and Scope, may be helpful to find new smells and to categorize them in the catalogue, in the sense that human analysts can look for new smells with these two orthogonal views clues.%

Also, the collected sample use case descriptions may not be enough regarding the generality.
Although we have tried to increase the variety of samples by including those of multiple domains, we cannot guarantee that the samples are the representative of the use case descriptions in general.

\subsubsection{Internal Validity}

We should explore the factors affecting the obtained results other than those of our approach.
One of the possibilities of these factors is bias at constructing the correct set of bad smell instances by our eight study participants.
However, their skills were sufficient and had experience in writing use case descriptions, and so we believe that the quality of the correct set was sufficient.
In addition, they performed their tasks independently and there are no effects among our study participants.

The second factor was the possibility of developing metrics arbitrarily.
However, we separated the examples of use case descriptions into two disjoint sets; one (30 examples) was for constructing the metrics and another (8 examples) for evaluating our detector.

The third factor was categorizing bad smells by one person when the bad smell catalogue was developed.
It leads to the possibility to wrong categorization.
Thus, the validation of the catalogue done by multiple persons is necessary as one of the future work.%

The fourth one is the problem domains whose use case descriptions we used.
We do not necessarily think that we could cover all problem domains.
Use case descriptions of a certain problem domain allows us to get the new entries of the catalogue and the different evaluation results.
Clarifying a role of problem domains is also one of the future work.%

The fifth one is that we did not have the agreement process of the correct set of smell instances by the study participants, and we used the union of the results of the participants.
Although one of the authors confirmed all of the results were valid except for one of them, one can disagree with smell instances produced by another.
Preparing smell instance oracle of high quality is also regarded as one of the future work.%

\section{Related Work}\label{s:relatedwork}

We focus on the studies to detect poor use case descriptions.
As mentioned in Section~\ref{sss:rq1-2}, we could find three popular checklists.
Phalp \etal \cite{phalp-sqj2007} developed the checklist called 7C's to measure the quality of use case descriptions.
It was based on an investigation about what quality was preferable for use case descriptions from the view of the understandability of natural language sentences.
T\"{o}rner \etal \cite{torner-isese2006} developed the 12 criteria and provided the questionnaires written in the natural language to decide if descriptions are bad smells or not.
These 12 criteria could be used to detect seven bad smells that T\"{o}rner \etal listed up.
Anda and Sj{\o}berg \cite{anda-seke2002} proposed the classification technique of flaws in use case models based on their locations and characteristics, and provided a checklist constructed from this classification.
The checklist consists of questionnaires written in natural language related to locations and characteristics.
The differences of these checklists to ours were discussed in Section~\ref{sss:rq1-2} from the view of the coverage of bad smells.
In addition, all of them are written in natural language, and the sufficient skills of their users are required.

Some researchers have adopted natural language techniques to analyze ambiguity and vagueness of natural language sentences as requirements descriptions.
Yang \etal discussed nocuous ambiguity such as the scope of conjunctives ``and'' and ``or'', and automated to detect its occurrences \cite{yang-re2010,yang-ase2010}.
They applied their approach to usual natural language sentences, not use case descriptions.
Use case descriptions are rather short sentences and may be incomplete ones.
Liu \etal \cite{liu-ase2014} tried to detect defects in use case documents automatically.
In this approach, a use case description is transformed into an activity diagram based on dependency parsing techniques, and defect detection is applied to the obtained activity diagram.
Fantechi \etal \cite{fantechi-rej2003} summarized the application of linguistic techniques to use case analysis.
Although their approach is complementary to ours, this line of approaches based on natural language processing is effective because it can extract a partial meaning of a use case description and can be utilized to increase the detection coverage of our smell catalogue.
Text2Test \cite{sinha-icst2010} is an integrated environment for authoring use cases with a model-based checking is available.
It is beneficial to develop a similar environment to realize Just-in-Time smell detection.

\emph{Refactorings}, transformations that solve bad smells, have been used mainly in source code level \cite{refactoring}, but the framework and terminology are not limited to the area related to source code.
The framework, i.e., detecting quality problems of software artifacts as smells and fixing them by refactoring to improve the quality with preserving their core aspects, is straightforward and is applicable to software artifacts in the scope of requirements engineering, such as use case models \cite{yu-ase2004,xu-apsec2004}, feature models \cite{vander-gpce2006}, goal models \cite{k_asano-mreba2017}, and natural language documents \cite{RefactoringHTML,aversano-icse2013}.
The goal of our approach can be classified as the same category; it regards the detection of problematic portions in use case documents as bad smells.
The techniques on refactoring use case models \cite{yu-ase2004,xu-apsec2004} focus on their structure improvement as model transformation, but not on the detection of bad smells.
Their approaches are complementary to ours.

\section{Conclusion}\label{s:conclusion}
This paper discussed the automated technique to detect the occurrences of poor parts in use case descriptions.
After clarifying bad smells~(symptoms) of poor descriptions from the two views: characteristics and scope, we defined a catalogue of bad smells for use case descriptions.
Furthermore, we applied the GQM approach to the bad smells and developed metrics to automate the detection of bad smells.
Finally, we could get the catalogue of 60 bad smells and an automated smell detector to indicate 24 bad smells of them.
Our analytic and experimental results showed that our direction is promising and some improvement is necessary.
As a result, we could automate to detect the bad smells of syntax-related issues only.
By capturing the catalogue with a bird's eye view, as the benefit of the catalogue, we can classify on the catalogue the bad smells that could be automated, and the others, which could not be done, are semantic-related issue.
Developing more sophisticated techniques for processing semantics of natural language is necessary to automate to detect them, and it is one of the future work.
However, we think that our catalogue can be used as a check list for human analysts to detect in industry.
The categorization of bad smells on the catalogue helped us to develop reusable metric:s and predicates for detecting the bad smells of the same category.
Same or similar metrics and predicates could be used to detect the bad smells of the same category and it allowed us to develop the tool without redundant efforts.

Future work can be summarized as follows:
\begin{itemize}
  \item
  More experimental analyses on use case models of wide varieties of problem domains and in a practical level.
  \item
  Related to the future work on more experimental analysis, collecting more use case descriptions with wide varieties of problems domains and with practical levels to improve our catalogues and metrics.
  We used a limited set of use case descriptions and it caused to be the possibility of the catalogues which could not covered all of bad smells yet.
  Accumulating various types of use case descriptions is useful to make the catalogue more comprehensive.
  In addition, we should investigate whether types of our bad smells are specific to problems domains or not and clarify the roles of the problems domains to the bad smells.
  \item
  Redoing categorization of bad smells by multiple persons and reviewing it to improve the quality of the catalogue.
  As discussed in Section~\ref{ss:threats}, when we developed the catalogue, only one person categorized the obtained bad smells.
  To improve the objectivity of the categorization, we should have an involvement of multiple persons to categorize the bad smells again and to review the catalogue.
  \item
  Developing a bad smell catalogue for use case diagrams and an automated tool to detect the bad smells.
  Combining them with the results of this work can be considered.
  \item
  Enhancing metrics, in particular for detecting bad smells related to semantics.
  Our current metrics are on for syntactic or surface features of use case descriptions.
  To get a more powerful detector, we need a deeper analysis of natural language sentences including semantic processing.
  In addition, since a use case model includes use case diagrams, the information that they have may be useful.
  \item
  Technique to support the improvement of detected bad smells, so-called \emph{refactoring} of use case models.
\end{itemize}

The list of use case descriptions that we used, the definition of all the smells, and the detailed comparison between catalogues are included in the appendix of this paper \cite{appendix}.

\section*{Acknowledgments}
This work was partly supported by JSPS Grants-in-Aids for Scientific Research (JP18K11237 and JP18K11238).

\IEEEtriggeratref{11}

\bibliographystyle{IEEEtran}

\begin{thebibliography}{10}
\providecommand{\url}[1]{#1}
\csname url@samestyle\endcsname
\providecommand{\newblock}{\relax}
\providecommand{\bibinfo}[2]{#2}
\providecommand{\BIBentrySTDinterwordspacing}{\spaceskip=0pt\relax}
\providecommand{\BIBentryALTinterwordstretchfactor}{4}
\providecommand{\BIBentryALTinterwordspacing}{\spaceskip=\fontdimen2\font plus
\BIBentryALTinterwordstretchfactor\fontdimen3\font minus
  \fontdimen4\font\relax}
\providecommand{\BIBforeignlanguage}[2]{{%
\expandafter\ifx\csname l@#1\endcsname\relax
\typeout{** WARNING: IEEEtran.bst: No hyphenation pattern has been}%
\typeout{** loaded for the language `#1'. Using the pattern for}%
\typeout{** the default language instead.}%
\else
\language=\csname l@#1\endcsname
\fi
#2}}
\providecommand{\BIBdecl}{\relax}
\BIBdecl

\bibitem{ApplyingUseCases}
S.~Geri and W.~Jason, \emph{Applying Use Cases: A Practical Guide}.\hskip 1em
  plus 0.5em minus 0.4em\relax Addison Wesley Longman, 1998.

\bibitem{el-attar-sosym2010}
M.~El-Attar and J.~Miller, ``Improving the quality of use case models using
  antipatterns,'' \emph{Software {\&} Systems Modeling}, vol.~9, no.~2, pp.
  141--160, 2010.

\bibitem{phalp-sqj2007}
K.~T. Phalp, J.~Vincent, and K.~Cox, ``Assessing the quality of use case
  descriptions,'' \emph{Software Quality Journal}, vol.~15, no.~1, pp. 69--97,
  2007.

\bibitem{torner-isese2006}
F.~T{\o}rner, M.~Ivarsson, F.~Pettersson, and P.~\"{O}hman, ``Defects in
  automotive use cases,'' in \emph{Proceedings of the 5th ACM/IEEE
  International Symposium on Empirical Software Engineering (ISESE 2006)},
  2006, pp. 115--123.

\bibitem{refactoring}
M.~Fowler, \emph{Refactoring: Improving the Design of Existing Code}.\hskip 1em
  plus 0.5em minus 0.4em\relax Addison Wesley, 1999.

\bibitem{GQM}
V.~Basili, C.~Caldiera, and D.~Rombach, ``Goal, question, metric paradigm,''
  \emph{Encyclopedia of Software Engineering}, vol.~1, pp. 528--532, 1994.

\bibitem{NEologd}
T.~H. Toshinori~Sato and M.~Okumura, ``Implementation of a word segmentation
  dictionary called mecab-ipadic-neologd and study on how to use it effectively
  for information retrieval (in {Japanese}),'' in \emph{Proceedings of the 23th
  Annual Meeting of the Association for Natural Language Processing (NLP
  2017)}, 2017, pp. NLP2017--B6--1.

\bibitem{anda-seke2002}
B.~Anda and D.~I.~K. Sj{\o}berg, ``Towards an inspection technique for use case
  models,'' in \emph{Proceedings of the 14th International Conference on
  Software Engineering and Knowledge Engineering (SEKE 2002)}, 2002, pp.
  127--134.

\bibitem{yang-re2010}
H.~Yang, A.~N. De~Roeck, V.~Gervasi, A.~Willis, and B.~Nuseibeh, ``Extending
  nocuous ambiguity analysis for anaphora in natural language requirements,''
  in \emph{Proceedings of the 18th {IEEE} International Requirements
  Engineering Conference ({RE'10})}, 2010, pp. 25--34.

\bibitem{yang-ase2010}
H.~Yang, A.~Willis, A.~N. De~Roeck, and B.~Nuseibeh, ``Automatic detection of
  nocuous coordination ambiguities in natural language requirements,'' in
  \emph{Proceedings of the 25th {IEEE/ACM} International Conference on
  Automated Software Engineering ({ASE 2010})}, 2010, pp. 53--62.

\bibitem{liu-ase2014}
S.~Liu, J.~Sun, Y.~Liu, Y.~Zhang, B.~Wadhwa, J.~S. Dong, and X.~Wang,
  ``Automatic early defects detection in use case documents,'' in
  \emph{Proceedings of the 29th ACM/IEEE International Conference on Automated
  Software Engineering (ASE 2014)}, 2014, pp. 785--790.

\bibitem{fantechi-rej2003}
A.~Fantechi, S.~Gnesi, G.~Lami, and A.~Maccari, ``Applications of linguistic
  techniques for use case analysis,'' \emph{Requirements Engineering}, vol.~8,
  no.~3, pp. 161--170, 2003.

\bibitem{sinha-icst2010}
A.~Sinha, S.~M.~S. Jr., and A.~Paradkar, ``{Text2Test}: Automated inspection of
  natural language use cases,'' in \emph{Proceedings of the 3rd International
  Conference on Software Testing, Verification and Validation (ICST 2010)},
  2010, pp. 155--164.

\bibitem{yu-ase2004}
W.~Yu, J.~Li, and G.~Butler, ``Refactoring use case models on episodes,'' in
  \emph{Proceedings of the 19th IEEE International Conference on Automated
  Software Engineering (ASE 2004)}, 2004, pp. 328--335.

\bibitem{xu-apsec2004}
J.~Xu, W.~Yu, K.~Rui, and G.~Butler, ``Use case refactoring: A tool and a case
  study,'' in \emph{Proceedings of the 11th Asia-Pacific Software Engineering
  Conference (APSEC 2004)}, 2004, pp. 484--491.

\bibitem{vander-gpce2006}
V.~Alves, R.~Gheyi, and T.~Massoni, ``Refactoring product lines,'' in
  \emph{Proceedings of the 5th international Conference on Generative
  Programming and Component Engineering (GPCE 2006)}, 2006, pp. 201--210.

\bibitem{k_asano-mreba2017}
K.~Asano, S.~Hayashi, and M.~Saeki, ``Detecting bad smells of refinement in
  goal-oriented requirements analysis,'' in \emph{Proceedings of the 4th
  International Workshop on Conceptual Modeling in Requirements and Business
  Analysis (MReBa 2017)}, 2017, pp. 122--132.

\bibitem{RefactoringHTML}
E.~R. Harold, \emph{Refactoring {HTML}: Improving the Design of Existing Web
  Applications}.\hskip 1em plus 0.5em minus 0.4em\relax Addison-Wesley, 2012.

\bibitem{aversano-icse2013}
L.~Aversano, G.~Canfora, G.~De~Ruvo, and M.~Tortorella, ``An approach for
  restructuring text content,'' in \emph{Proceedings of the 35th International
  Conference on Software Engineering (ICSE 2013)}, 2013, pp. 1225--1228.

\bibitem{appendix}
Y.~Seki, S.~Hayashi, and M.~Saeki, ``Appendix of ``{D}etecting bad smells in
  use case descriptions'','' Zenodo, 2019,
  \url{https://doi.org/10.5281/zenodo.3344974}.

\end{thebibliography}

\includepdf[pages=-]{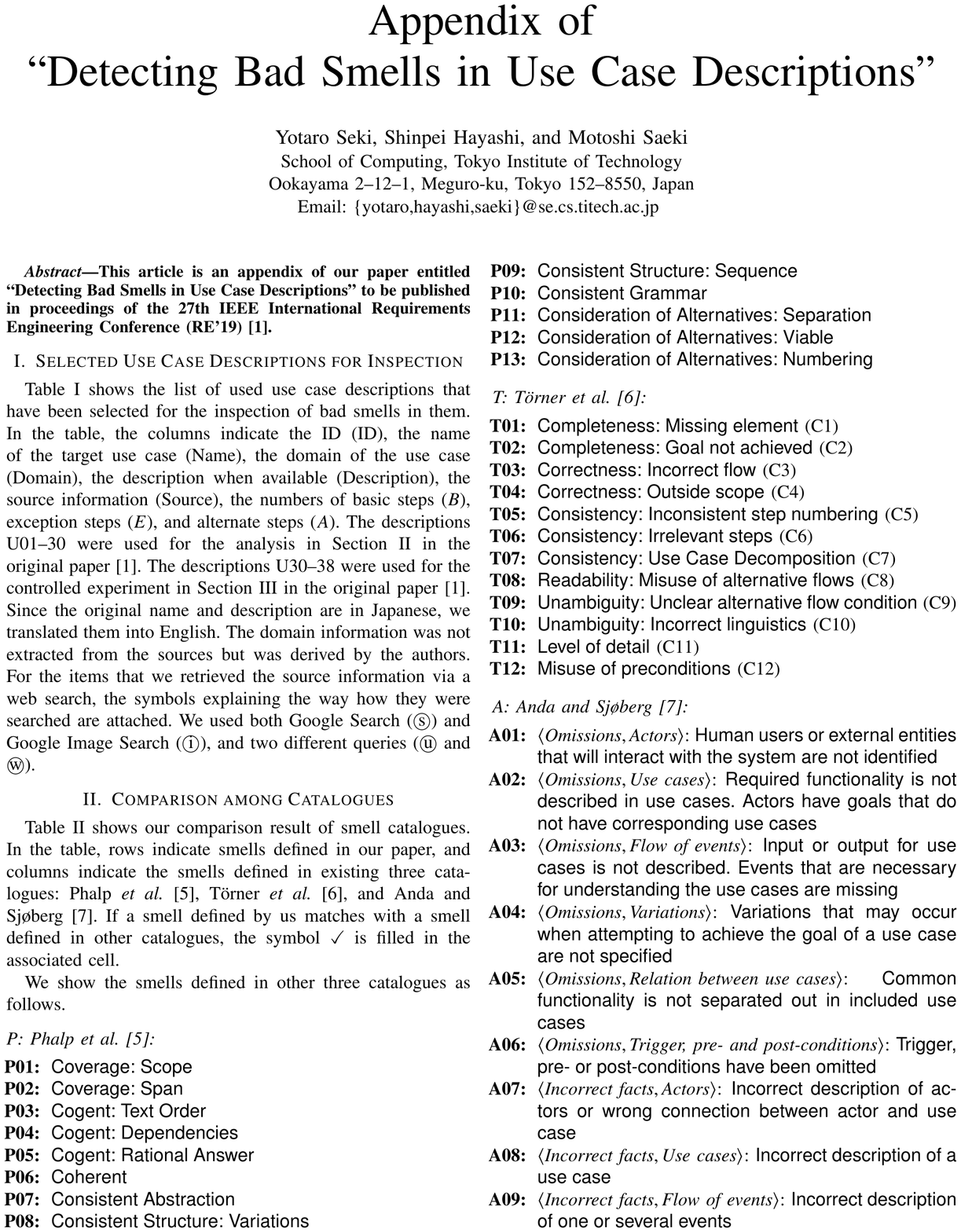}

\end{document}